\pdfoutput=1
\documentclass[11pt]{article}

\usepackage[margin=1in]{geometry}

\usepackage[pdftex]{graphicx}

\usepackage{amsmath,amssymb,amsthm}
\usepackage{float}
\usepackage{algorithm}
\usepackage{algpseudocode}
\usepackage{wrapfig}
\usepackage[utf8]{inputenc}
\usepackage[T1]{fontenc}
\usepackage{hyperref}
\usepackage{etoc}
\hypersetup{
  pdftitle={Symplectic Reck: In-Situ Learning of Gaussian Quantum Operations},
  pdfauthor={Janet F. Zhong et al.}
}
\usepackage{url}
\usepackage{booktabs}
\usepackage{amsfonts}
\usepackage{nicefrac}
\usepackage[activate={true,nocompatibility},expansion=false]{microtype}
\usepackage{xcolor}

\newtheorem{theorem}{Theorem}

\usepackage[numbers,sort&compress]{natbib}

\usepackage{etoolbox}
\appto\maketitle{%
  \begingroup
  \footnotetext{40th Conference on Neural Information Processing Systems (NeurIPS 2026).}%
  \endgroup
}
\title{Symplectic Reck: In-Situ Learning of Gaussian Quantum Operations}

\author{
  Janet F. Zhong$^{1}$ \quad
  Renwen Yu$^{1,2}$ \quad
  Paul-Alexis Mor$^{3,4}$ \quad
  Charles Roques-Carmes$^{3}$ \\
  Aviv Karnieli$^{5}$\quad
  David A. B. Miller$^{1}$ \quad
  Shanhui Fan$^{1}$\\[0.5em]
  \small $^{1}$Edward L. Ginzton Laboratory, Stanford University, Stanford, CA, USA \\
\small $^{2}$CNRS, \'Ecole Centrale de Lyon, INSA Lyon, Universit\'e Claude Bernard Lyon 1, \\
  \small CPE Lyon, INL, UMR 5270, F-69134, Ecully, France \\
  \small $^{3}$Institute of Science and Technology Austria, Klosterneuburg, Austria \\
  \small $^{4}$Corps des Mines, Mines Paris, Universit\'e PSL, Paris, France \\
  \small $^{5}$Technion -- Israel Institute of Technology, Haifa, Israel
}

\date{}

\newenvironment{ack}{\section*{Acknowledgments}}{}

\begin{document}\etocdepthtag.toc{main}\maketitle

\begin{abstract}
Programmable photonic circuits are a natural platform for physical learning, where hardware parameters are trained directly from measured feedback. We propose a physical learning algorithm and architecture for continuous-variable quantum systems where a photonic circuit is trained in situ to invert an unknown lossless Gaussian quantum operation. We prove that a generic symplectic matrix $S\in\mathrm{Sp}(2N,\mathbb{R})$
admits a decomposition into a triangular array of two-mode symplectic gates
on neighboring modes, followed by single-mode symplectic gates. We then
introduce the symplectic Reck mesh, which physically realizes this
factorization, generalizing the unitary Reck mesh widely used in optical
neural networks. When placed after an unknown Gaussian unitary, the mesh is trained in situ by nulling local response blocks one at a time using optical probes and measurements. We identify a local condition number that quantifies the squeezing required for a nulling step and its sensitivity to measurement error. We numerically demonstrate finite-shot training and robustness to systematic gate-calibration errors. Our work extends self-configuring optical networks from unitary to symplectic transformations, bringing physical learning to continuous-variable quantum photonic hardware.\footnote{Code is available at \url{https://github.com/janetzhong/symplectic-reck}.}
\end{abstract}

\section{Introduction}
\label{sec:intro}

Physical learning trains a physical system directly from measured responses,
rather than learning a digital model and compiling it onto the hardware
afterward~\cite{stern2021supervised,dillavou2022demonstration,wright2022deep,momeni2025training}.
Training from measured responses can compensate for fixed calibration
errors. In photonics, an example is
\emph{self-configuring optics}~\cite{miller2013self}, where a programmable
optical mesh is tuned using measurements of its response to optical probes.
Self-configuring photonic circuits have been developed for undoing unwanted
mode mixing~\cite{annoni2017unscrambling,milanizadeh2022separating},
finding optimal communication channels~\cite{seyedinnavadeh2023determining},
spectral filtering~\cite{miller2025universal,valdez2026programmable},
and variational optical
diagonalization~\cite{roquescarmes2024measuring,mor2026separating,karnieli2025variational,roquescarmes2025automated}.
Optical inversion or mode unscrambling can be useful for recovering fields, images, or
communication channels after a multimode system has mixed them~\cite{butaite2022how,kupianskyi2024all}.

Many self-configuring meshes are based on the unitary Reck architecture,
which realizes an arbitrary unitary transformation using a triangular
array of two-mode unitary gates together with single-mode phase
shifts~\cite{reck1994experimental,miller2013self,perez2020principles}.
These circuits are built from phase shifters and beam splitters and perform passive (photon-number-conserving) linear optics.
Continuous-variable quantum optics also involves squeezing, which reduces quantum uncertainty in one field quadrature while increasing it in the conjugate quadrature. Squeezing is an active (photon-number-nonconserving) operation.
Both passive mode mixing and squeezing belong to the class of lossless
Gaussian operations~\cite{adesso2014continuous,serafini2017quantum,weedbrook2012gaussian,fabre2020modes}. The linear part of a lossless Gaussian operation is described by a real symplectic matrix. The unitary Reck mesh cannot implement general symplectic transformations. Extending self-configuration to symplectic transformations would allow an optical mesh to implement or reverse both mode mixing and squeezing. This could help correct unwanted mode coupling in continuous-variable quantum communication and squeezed-light processing~\cite{filip2005elimination,kovalenko2021cross,ren2022topologically}.

In this paper, we develop a method for learning the inverse of an unknown lossless Gaussian operation directly in hardware. A conventional approach would be to first characterize the unknown transformation using tomography, where one sends optical probes to estimate the full symplectic matrix of the lossless Gaussian operation~\cite{teo2021highly,wang2013efficient,kumar2020optimal,grove2025characterizing,fanizza2025efficient}. From this, one can compute the inverse of the measured symplectic matrix digitally and use this to decompose the inverse into available optical gates. One would then need to compile the resulting parameters into the hardware~\cite{bloch1962canonical,cariolaro2016bloch,kalajdzievski2021exact,volkoff2021universal}. This route is natural when the hardware parameters are accurately calibrated. In practice, this is rarely the case as fabrication variations, unintended phase shifts, heat transfer between nearby components (thermal crosstalk), inaccurately calibrated squeezing strengths and more can cause implemented gates to differ from their intended settings~\cite{alexiev2021calibrating,fyrillas2024scalable,zheng2024global}. An in-situ method instead programs the inverse from feedback measurements of the system. This allows for the training of the inverse to account for static, repeatable implementation errors, similar to other self-configuring photonic
meshes~\cite{miller2015perfect}.

We introduce the \emph{symplectic Reck mesh}, a triangular array of
two-mode symplectic gates followed by a layer of single-mode symplectic gates. Placed after an unknown device $S_{\rm unk}$, the mesh, with transformation
$S_\theta$, is trained so that $M=S_\theta S_{\rm unk}\approx I$. The training is performed by sequentially nulling $2\times2$ quadrature-response blocks,
setting one gate at a time from local measurements of the system. Gate settings are obtained either by optimization using
measured feedback or by a closed-form update calculated from optical probe
measurements. Both approaches use fresh measurements as training proceeds,
so later steps can partly compensate for errors in earlier gates.

\paragraph{Contributions.}
\begin{enumerate}\setlength{\itemsep}{1pt}\setlength{\topsep}{2pt}
    \item \textbf{Physical learning task.} We formulate in-situ inversion of an
    unknown lossless Gaussian operation as a physical learning problem. The
    learned object is the trained circuit itself rather than a digital model
    of $S_{\rm unk}$.

    \item \textbf{Theory and architecture.} We show that the Reck factorization
    extends to generic matrices in $\mathrm{Sp}(2N,\mathbb{R})$ through
    sequential two-mode block nulling on neighboring modes and a final layer of
    single-mode gates. We identify exactly when each local nulling step is
    possible and derive a condition number $\kappa$ that quantifies how
    difficult it is. We propose a symplectic Reck mesh that realizes this
    factorization in hardware.

    \item \textbf{Local training algorithm.} We show how to carry out this
    factorization using only measured responses of the physical system. Each
    gate is set either by feedback optimization of a measured loss or by a
    closed-form update from response blocks estimated with optical probes and
    measurements.

    \item \textbf{Experiments.} We numerically show that the condition number
    $\kappa$ tracks trends in residual error as we vary the mode number, as the
    target approaches a nulling failure, and as we vary the squeezing strength
    of the target matrix. We evaluate the effects of finite-shot noise and
    compare measured closed-form updates, gradient-based feedback, and
    gradient-free feedback up to $N=30$. We also show that in-situ nulling can
    outperform tomography followed by offline compilation in the presence of
    gate-calibration errors.
\end{enumerate}

\section{Background and problem setup}
\label{sec:setup}
We first review the continuous-variable concepts used in this paper and
then define the in-situ inversion task.
\paragraph{Quadratures and Gaussian operations.}
Each optical mode $m$ is described by two quadratures, $x_m$ and $p_m$,
which correspond to the real and imaginary parts of the field amplitude. For $N$
modes we can define the quadrature vector $r=(x_1,p_1,\ldots,x_N,p_N)^\top$.
A Gaussian operation transforms the mean quadratures as $\langle r_{\rm out}\rangle=S\langle r_{\rm in}\rangle+d$, where $S$ is a real $2N\times2N$ matrix and $d$ is a fixed displacement
vector that shifts the mean quadratures. Not every matrix $S$ is physically allowed. The set of physically allowable $S$ are those that preserve the canonical commutation relations. This holds exactly when $S^\top J_{2N}S=J_{2N}$, where
$J_{2N}=\bigoplus_{m=1}^{N}J_2$ is the symplectic form and
$J_2=\left(\begin{smallmatrix}0&1\\-1&0\end{smallmatrix}\right)$.
Matrices satisfying this condition
are called \emph{symplectic} and form
the group $\operatorname{Sp}(2N,\mathbb R)$~\cite{adesso2014continuous,serafini2017quantum,weedbrook2012gaussian,fabre2020modes}. Symplectic matrices can be built from three kinds of elementary optical
components. Phase shifters rotate the quadratures of individual modes, beam splitters mix quadratures between modes and squeezers increase the
uncertainty in one quadrature while decreasing it by the same factor in the conjugate quadrature. Phase shifters and beam splitters are orthogonal and symplectic (orthosymplectic) in quadrature space and squeezers are symplectic but not orthogonal. 

\paragraph{Task and access model.}
We consider an unknown lossless Gaussian operation $S_{\rm unk}\in\operatorname{Sp}(2N,\mathbb R)$,
with unknown displacement $d$, followed by a programmable mesh $S_\theta$. Our goal is to choose mesh settings in $S_\theta$ so that the combined transformation
$M=S_\theta S_{\rm unk}\approx I$, and then to cancel the remaining
displacement. The learned $S_\theta$ performs the inverse of $S_{\rm unk}$ up to measurement and implementation errors. We assume the unknown device stays
fixed during training.  To learn the settings, we use optical probes through the combined system and
measure its quadrature
response~\cite{lvovsky2009continuous,weedbrook2012gaussian,fabre2020modes}. The gates that have not been set are set to identity. We assume the response is measured directly after the gate that is being set.  This is similar to the local monitors used in unitary
meshes~\cite{annoni2017unscrambling,milanizadeh2022separating}.

\section{Related work}
\label{sec:related}

\paragraph{Self-configuring photonics and in-situ inversion.}
Self-configuring photonics uses measured feedback to tune an optical circuit
directly~\cite{miller2013self}. Sequential in-situ nulling with two-mode unitary gates has been
demonstrated experimentally to undo unwanted mode mixing and separate
optical beams~\cite{annoni2017unscrambling,milanizadeh2022separating}. These tasks
are closely related to the unitary analogue of our in-situ inverse task,
$U_\theta U_{\rm unk}\approx I$, but may leave residual output phase shifts,
which do not affect channel separation. Optically inverting the mode mixing in a multimode fiber can enable
imaging through it, with potential applications in optical
communications and computing~\cite{butaite2022how,kupianskyi2024all}. Removing unwanted mode mixing is also relevant to continuous-variable
quantum communication and squeezed-light
processing~\cite{filip2005elimination,kovalenko2021cross,ren2022topologically}. Our task extends in-situ inversion from $\mathrm U(N)$ to
$\operatorname{Sp}(2N,\mathbb R)$, allowing squeezing.

\paragraph{Learning Gaussian transformations and compiling inverses.}
A conventional approach to inverting an unknown optical operation is to
reconstruct the transformation using
tomography~\cite{teo2021highly,wang2013efficient,kumar2020optimal,grove2025characterizing,fanizza2025efficient},
compute its inverse, decompose it into available gates, and program those gates
into the hardware~\cite{bloch1962canonical,cariolaro2016bloch,kalajdzievski2021exact,volkoff2021universal}.
The final step can be affected by calibration errors, including fabrication
variations, phase offsets, thermal crosstalk, and squeezer
miscalibration~\cite{alexiev2021calibrating,fyrillas2024scalable,zheng2024global}. We typically need detailed models of how the hardware responds to control settings~\cite{fyrillas2024scalable,zheng2024global,miilanizadeh2020control} to account for such errors. Self-configuring algorithms use feedback to adjust circuit parameters, which can avoid the need for detailed calibration models and simplify experimental methods.

\paragraph{Symplectic QR-like factorizations.}
We derive a symplectic QR-like factorization that generalizes the Reck
construction~\cite{reck1994experimental,miller2013self}, decomposing a generic matrix in $\mathrm{Sp}(2N,\mathbb{R})$
into two-mode $\mathrm{Sp}(4,\mathbb{R})$ gates on neighboring modes and
single-mode $\mathrm{Sp}(2,\mathbb{R})$ gates. Similar algorithms for the
symplectic QR factorization, using local symplectic transformations, were
derived earlier in
Refs.~\cite{singer2003rounding,singer2012orthosymmetric}. Here, we give a
different construction of each nulling step, prove that our nulling
sequence succeeds for generic symplectic matrices, and show that the
factorization can be realized and trained in hardware. Other symplectic
QR-like and SR decompositions exist but differ from ours in their objectives
and resulting matrix
forms~\cite{benner1998symplectic,bunseGerstner1986matrix,salam2008optimal,sadkane2008note}.

\section{The symplectic Reck mesh}
\label{sec:architecture}
We now describe the mesh architecture, its gate training order, and how
each gate is set from measured responses.

\begin{figure}[t]
    \centering
    \includegraphics[width=\linewidth]{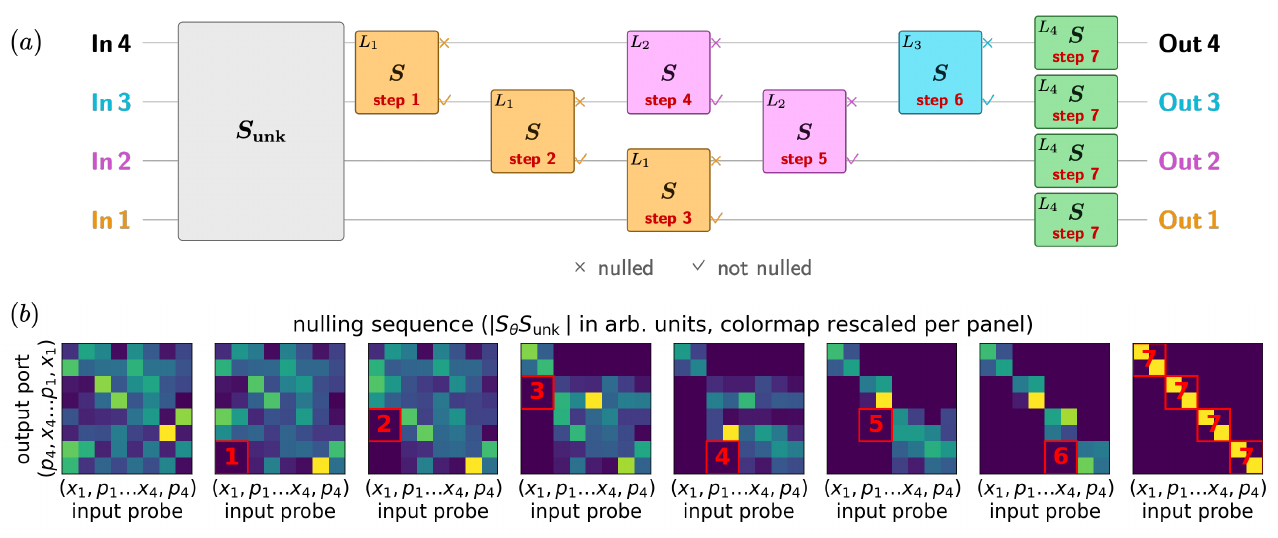}
    \caption{(a) Symplectic Reck mesh for $N=4$ modes. Red step numbers show the gate
    training order. The final single-mode gates can be trained in parallel.
    (b) Simulated response $|S_\theta S_{\rm unk}|$ during the training steps. Red boxes mark the blocks being
    nulled and correspond to the step numbers in (a). Two-mode gates bring the response close to block diagonal,
    and single-mode gates bring it close to the identity, with a final
    residual of $1.1\times10^{-4}$. Each panel uses its own color scale.}
    \label{fig:circuit}
\end{figure}

\paragraph{Block notation.}
Because we use the quadrature ordering
$r=(x_1,p_1,\ldots,x_N,p_N)^\top$ which is grouped by mode, we can view the $2N\times2N$ response $M$
as an $N\times N$ grid of $2\times2$ blocks, with one block row and one block
column per mode. Block
$M_{j,k}=\left(\begin{smallmatrix}M_{x_j,x_k}&M_{x_j,p_k}\\M_{p_j,x_k}&M_{p_j,p_k}\end{smallmatrix}\right)$
describes how the output quadratures of mode $j$ respond to the input
quadratures of mode $k$. Here, $M_{x_j,p_k}$ is the response of output
$x_j$ to input $p_k$.

\paragraph{Architecture.}
The mesh contains a triangular array of $N(N-1)/2$ two-mode symplectic gates acting on neighboring modes, followed by one single-mode symplectic gate per mode
(Fig.~\ref{fig:circuit}). Each two-mode gate can implement any element of
$\operatorname{Sp}(4,\mathbb R)$. In hardware, it can be built from two passive
two-mode interferometers with two single-mode squeezers between them
(a Bloch--Messiah decomposition~\cite{bloch1962canonical,cariolaro2016bloch,fabre2020modes}).
Each single-mode gate is a squeezer between two phase shifters.
Appendix~\ref{SM:operations} gives explicit parameterizations and illustrates
these implementations in Fig.~\ref{fig:gate-bm}.

\paragraph{Gate training order.}
We null quadrature-response blocks of $M$ one gate at a time,
leaving each gate fixed after training. The nulling order goes through  block columns
$k=1,\ldots,N-1$, nulling the blocks below the diagonal from bottom to
top. The gates used for column $k$ form layer $L_k$ in the triangular array. A two-mode gate acts on
modes $(j-1,j)$ to null $M_{j,k}$, in the order $j=N,N-1,\ldots,k+1$. With exact nulling, this order preserves the zeros in earlier columns. Once all blocks below the diagonal are zero, symplecticity forces the blocks above the diagonal to also be zero (Appendix~\ref{app:preservation}). The
single-mode gates then invert the remaining diagonal blocks. We denote the measured
estimate of each response block by $\widehat{M}_{j,k}$.
Algorithm~\ref{alg:symplectic-reck} summarizes the procedure and Fig.~\ref{fig:circuit} depicts it for four modes. Any remaining
displacement can be removed by applying an opposite displacement to each
output mode (not shown in Fig.~\ref{fig:circuit})~\cite{gerry2023introductory}.

\begin{algorithm}[t]
\caption{In-situ training of the symplectic Reck mesh. Gate settings are
obtained using either measured feedback or closed-form updates from
measured response blocks.}\label{alg:symplectic-reck}
\begin{algorithmic}[1]
\State Set every gate to the identity, so $S_\theta=I$.
\For{$k = 1, \ldots, N-1$} \Comment{block columns, left to right}
    \For{$j = N, \ldots, k+1$} \Comment{blocks in column $k$, bottom to top}
        \State Probe input mode $k$ and estimate $\widehat{M}_{j,k}$ (and $\widehat{M}_{j-1,k}$ for closed-form training).
        \State Set the two-mode gate on modes $(j-1,j)$ so that $\widehat M_{j,k}\approx0$.
    \EndFor
\EndFor
\For{$m = 1, \ldots, N$} \Comment{blocks along diagonal, parallelizable}
    \State Set the single-mode gate on mode $m$ so that $\widehat{M}_{m,m}\approx I_2$.
\EndFor
\State Measure the output mean with no input displacement and cancel it.
\State \Return the trained mesh $S_\theta$, with $S_\theta S_{\rm unk}\approx I$.
\end{algorithmic}
\end{algorithm}

\paragraph{Measuring a response block.}
We use coherent optical pulses (essentially laser pulses) as probes. To
estimate $M_{x_j,p_k}$, we set the mean input $p_k$ to $+\eta$ and $-\eta$
and measure the mean output $x_j$, giving
$\widehat M_{x_j,p_k}=(\langle x_j\rangle_{+\eta p_k}-\langle x_j\rangle_{-\eta p_k})/(2\eta)$.
Taking the difference cancels the unknown displacement. The other three
entries of $M_{j,k}$ are measured in the same way. Shifting either $x_k$ or
$p_k$ by $\pm\eta$ gives four inputs. Measuring $x_j$ and $p_j$ separately
for each input using homodyne detection gives eight measurement settings.
Heterodyne detection instead measures both output quadratures
together~\cite{lvovsky2009continuous,weedbrook2012gaussian,fabre2020modes}.

\paragraph{Measured closed-form or feedback-based updates.}
The algorithm can use either feedback training or measured closed-form
updates. In \emph{feedback} training, an optimizer adjusts the gate's
controls to minimize a measured loss: $\|\widehat M_{j,k}\|_F^2$ for a
two-mode gate, and $\|\widehat M_{m,m}-I_2\|_F^2$ for a single-mode gate in
the final layer. This does not require a calibration model of the gate, only
the ability to change its controls and observe the result. In
\emph{closed-form} training, we measure both blocks in the two active rows,
$\widehat M_{j-1,k}$ and $\widehat M_{j,k}$, and compute the nulling gate
directly from them (Section~\ref{sec:theory}). For the final layer, we
instead invert each measured diagonal block $\widehat M_{m,m}$ after
projecting it onto a symplectic matrix
(Appendix~\ref{app:experimental-setup}). This avoids iterative optimization
but requires a calibration model to set the current gate. In both methods,
each update uses only local $2\times2$ blocks, independent of $N$. The
effects of earlier gates are measured rather than predicted from a model, so
later gates can partly compensate for their errors
(Section~\ref{sec:experiments}). For fixed shot and evaluation budgets per
gate, training uses $O(N^2)$ measurement shots in total
(Appendix~\ref{app:training-cost}).

\section{Theory}
\label{sec:theory}
We now establish when the symplectic Reck mesh succeeds. We first
determine when a two-mode gate can null a response block, then prove that the full sequence succeeds for
generic symplectic matrices. We also derive a local condition number
that quantifies how difficult each nulling step is.

\subsection{When is a nulling step possible?}
\label{subsec:local-pivot}

Consider a step that aims to set the block $M_{j,k}$ to zero. We apply a gate
to output modes $j-1$ and $j$, which mixes the corresponding two block
rows of $M$. In column $k$, let
\begin{equation}
P=M_{j-1,k},
\qquad
E=M_{j,k},
\label{eq:local-window-main}
\end{equation}
where $E$ is the block we want to eliminate and $P$ is the block
directly above it. We seek a two-mode symplectic gate $G$ such that
\begin{equation}
G\begin{pmatrix}P\\E\end{pmatrix}
=\begin{pmatrix}Q\\0\end{pmatrix}.
\label{eq:local-nulling-main}
\end{equation}
Thus, $G$ replaces $E$ with zero while allowing $P$ to change to a new
block $Q$.

In the unitary Reck mesh, a beam splitter and phase shifters can always make
one of the two output responses zero. For a general symplectic response, this
is not always possible. To describe when it fails, we define
\begin{equation}
\Delta=\det P+\det E,
\label{eq:delta-main}
\end{equation}
and call it the \emph{pivot}, as it plays a similar role to pivots in
Gaussian elimination. Because symplectic gates preserve the symplectic form,
$\Delta$ is unchanged by any two-mode symplectic gate
(Appendix~\ref{app:genericitygate}). After nulling, the blocks become $Q$
and $0$, so $\det Q=\Delta$.

\paragraph{Local nulling criterion.}
If the two columns of $\left(\begin{smallmatrix}P\\E\end{smallmatrix}\right)$
are linearly independent, nulling is possible if and only if
$\Delta\neq0$. To see why a zero pivot\footnote{Geometrically, a zero pivot with
independent columns means that the columns span an isotropic subspace, on
which the symplectic form vanishes~\cite{daSilva2008lectures}.} prevents nulling, recall that
a symplectic gate is invertible and therefore preserves linear
independence. After nulling, the lower block is zero, so the columns
of $Q$ must remain linearly independent. Hence, $Q$ must be invertible.
This is incompatible with $\det Q=\Delta=0$.
Conversely, when $\Delta\neq0$, the two response columns can be
mapped into a single mode by a symplectic gate, as shown by the
explicit construction in Appendix~\ref{app:genericitygate}.
Similar nulling criteria were derived in
Refs.~\cite{singer2012orthosymmetric,langer2026matchgate} using different
explicit gates, assuming invertible $P$ or invertible $P$ and $E$,
respectively. If the two columns are linearly dependent, there is at most one
independent response to move into the upper mode. Nulling in this case is
always possible using only a beam splitter and phase shifters. Thus, the only obstruction is a zero pivot with two independent
response columns. Appendix~\ref{app:genericity} gives the full proof.

\subsection{When does the full sequence succeed?}
\label{subsec:global-factorization}

A nonzero pivot means one nulling step succeeds. However, each gate changes the blocks used in later steps.
We must therefore show that the full sequence does not
generically hit a zero pivot.

\begin{theorem}[Symplectic Reck decomposition]
\label{thm:generic}
For a generic symplectic matrix $S_{\rm unk}\in\operatorname{Sp}(2N,\mathbb R)$, the nulling sequence of Algorithm~\ref{alg:symplectic-reck} succeeds, giving $S_\theta S_{\rm unk}=I$. Equivalently, a generic symplectic matrix can be implemented by a triangular mesh of two-mode symplectic gates on neighboring modes, followed by one single-mode symplectic gate on each mode.
\end{theorem}

Here, generic means that the exceptions form a set of measure zero, so a
randomly chosen target fails with probability zero. We give an outline of
the proof. We show that the choice of gate at each step does not affect
later pivots. This lets us follow an explicit gate construction, given in
Appendix~\ref{app:genericitygate}, which turns every pivot into a ratio of
polynomials in the entries of $S_{\rm unk}$. A target can fail only if one
of these polynomials is zero. To show that they are not all identically
zero, we construct a passive target from $50{:}50$ beam splitters for
which every pivot is strictly positive
(Appendix~\ref{app:generic-proof}). The failing targets therefore form a
set of measure zero. Since a mesh that inverts $S_{\rm unk}^{-1}$ equals
$S_{\rm unk}$, generic symplectic matrices also admit a symplectic Reck
factorization. 

The symplectic Reck mesh does not invert every symplectic matrix, but a
modified version can. Appendix~\ref{app:universal} gives one such
   construction, which adds passive gates before each nulling step to avoid
   zero pivots, similar to the passive rotations in the symplectic QR
   algorithm of Ref.~\cite{singer2003rounding}.

\subsection{Conditioning: how difficult is a nulling step?}
\label{subsec:conditioning-noise}

The nonzero pivot condition above guarantees a nulling gate exists, but it does not
say how much squeezing is needed or how sensitive the nulling step is
to measurement errors. A small $|\Delta|$ may indicate that the nulling step is close to a failure
point, but it can also mean that the whole response is small.

 To distinguish these cases, we compare $|\Delta|$ with the overall size of
the response, measured by the Euclidean area spanned by its two columns. Let
$V=\left(\begin{smallmatrix}P\\E\end{smallmatrix}\right)$. For linearly
independent response columns, we define the local condition number
\begin{equation}
\kappa=\frac{\sqrt{\det(V^\top V)}}{|\Delta|}\geq1.
\label{eq:kappa}
\end{equation}
The columns of $V$ are the responses to inputs $x_k$ and $p_k$, and the
numerator is the ordinary area of the parallelogram they span. The pivot
$\Delta$ can be thought of as a symplectic area. When the columns are
projected onto the $(x,p)$ plane of each mode, they form parallelograms with
signed areas $\det P$ and $\det E$, and their sum is $\Delta$. Symplectic gates
preserve this sum, although the individual areas can change. After nulling, both columns lie in one mode, so the ordinary area equals
$|\det Q|=|\Delta|$. The condition number $\kappa$ is the ratio of the ordinary
area before nulling to this area after nulling. Because both
areas scale in the same way, $\kappa$ is also independent of the overall scale
of the response, whereas $|\Delta|$ alone is not.

Passive gates preserve both areas, so when $\kappa=1$, the nulling step can be
achieved using beam splitters and phase shifters with no squeezers. A large
$\kappa$ means that $|\Delta|$ is small relative to the area spanned by the
response columns. Nulling then requires strong squeezing and can amplify
measurement errors. See Appendix~\ref{app:minimal-gate} for more details.

\paragraph{Minimum-squeezing gate.}
It is possible to state the minimum squeezing needed for a nulling step. For any
gate that completes the nulling step, at least one of the two single-mode
squeezers in its Bloch--Messiah decomposition must have squeezing strength at
least $r_{\min}=\tfrac12\operatorname{arccosh}\kappa$. A passive two-mode
transformation (built from beam splitters and phase shifters) followed by a
two-mode squeezer of strength $r_{\min}$ can achieve this bound
(Appendix~\ref{app:minimal-gate}). We call this the
\emph{minimum-squeezing gate} and use it in all closed-form simulations in
Section~\ref{sec:experiments}. It has only four control parameters, so a
symplectic Reck mesh built from these gates has $N(2N+1)$ parameters, which matches
$\dim(\operatorname{Sp}(2N,\mathbb R))$ (Appendix~\ref{app:four-parameters}).

\paragraph{Noise sensitivity and sample complexity.}
The condition number $\kappa$ also measures how strongly a nulling step
amplifies noise. If a measured block differs from the true response by
$\delta$ (in Frobenius norm), the squeezing used for nulling amplifies this
error, and the minimum-squeezing gate leaves a residual of at most
$\sqrt{\widehat\kappa}\,\delta$ in the block it was meant to null, where
$\widehat\kappa$ is the condition number of the measured blocks
(Appendix~\ref{app:error-amplification}). Averaging over $q$ shots per
setting reduces $\delta$ as $1/\sqrt q$, so the residual after one nulling step
scales as $\sqrt{\kappa/q}$. Since $\kappa$ depends only on $P$ and $E$,
it can be estimated from the measured responses and monitored during
training. We report its largest value over all nulling steps,
$\kappa_{\max}$. The closed-form method already measures both blocks,
whereas feedback training would need an extra measurement of $P$.
By combining the nulling-step bound over the full sequence, we can derive a conditional
probably approximately correct (PAC) guarantee for closed-form training.
Under the conditions stated in Appendix~\ref{app:pac}, a polynomial number
of shots suffices to reach a target residual with high probability.

\section{Numerical experiments}
\label{sec:experiments}

\begin{figure}[t]
\centering
\includegraphics[width=\linewidth]{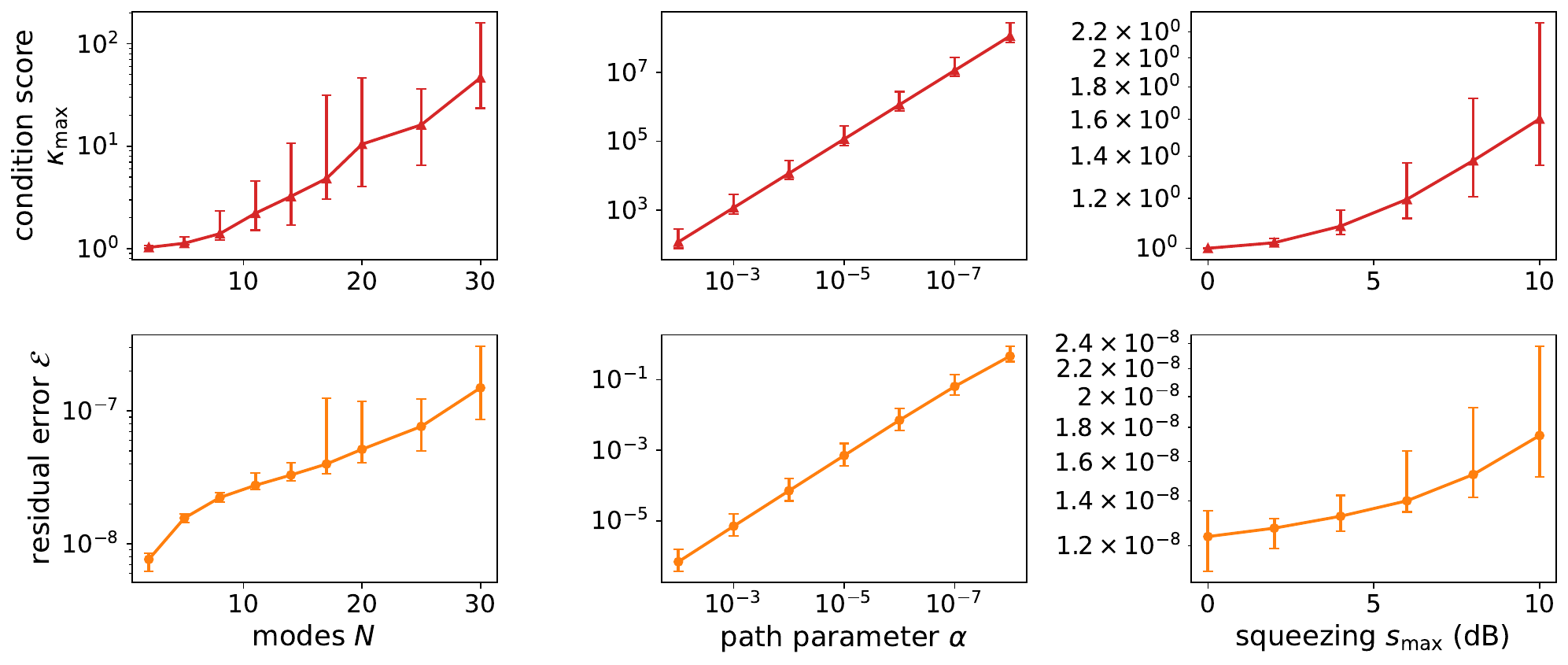}
\caption{The condition number tracks how much the mesh amplifies small errors.
Top row: $\kappa_{\max}$, the largest condition number along the nulling sequence.
Bottom row: final residual when each measured block has a small random error of size
$\epsilon=10^{-8}$. Left: increasing the number of modes $N$. Middle: moving the target matrix toward one where nulling fails
(smaller $\alpha$ is closer). Right: increasing the maximum target squeezing
$s_{\max}$ for $N=4$. The residual trend is similar to the $\kappa_{\max}$ trend in all three cases.
Markers show medians, and error bars show the 25th--75th percentiles over 60
trials (left) or 30 trials (middle and right). See Appendix~\ref{app:experimental-setup} for full experimental details.}
\label{fig:nulling_conditioning}
\end{figure}

\begin{figure}[t]
    \centering
    \includegraphics[width=0.9\linewidth]{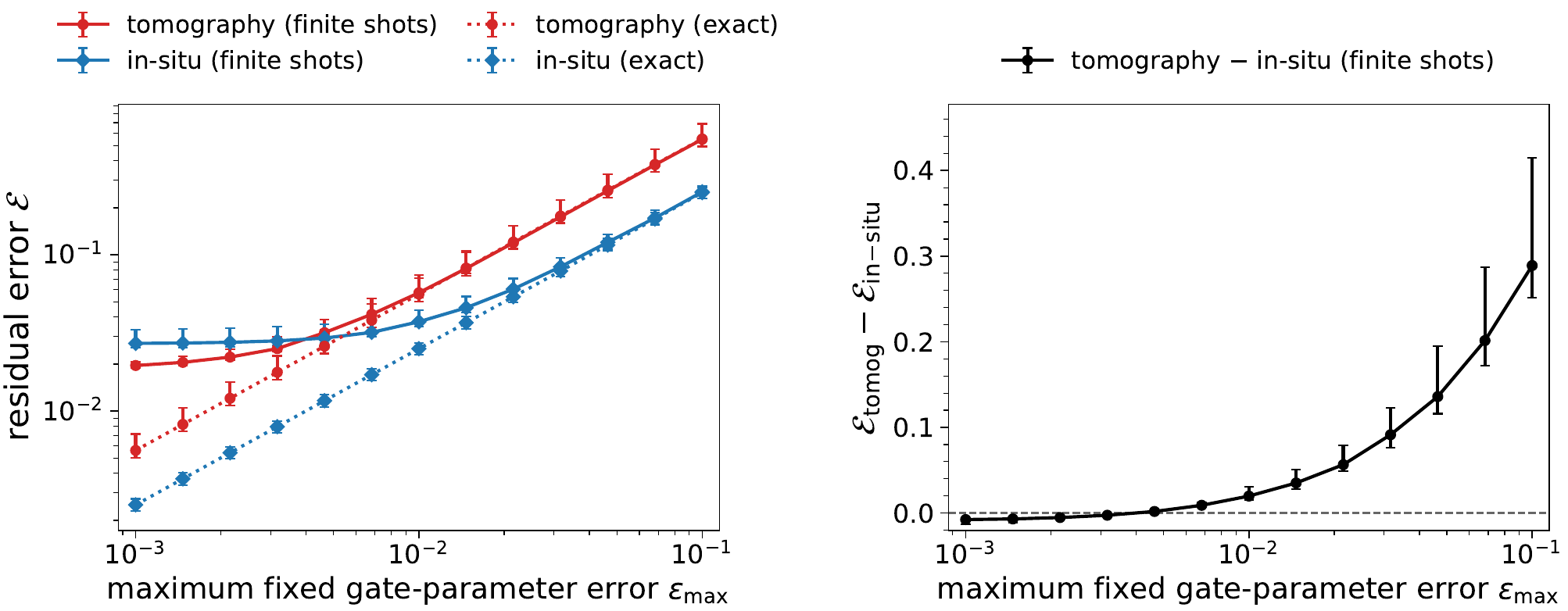}
\caption{In-situ training is more accurate than tomography followed by
offline compilation once fixed gate errors dominate measurement noise
($N=10$). Each gate is followed by a small symplectic error whose
parameters are drawn from
$\operatorname{Uniform}[-\epsilon_{\max},\epsilon_{\max}]$ and held fixed.
Left: final residual $\mathcal E$. Solid curves use $q=10^4$ shots per
homodyne setting, with the same total budget of $8N^2q$ samples for both
methods. Dotted curves use noiseless measurements with the same gate errors.
Right: $\mathcal E_{\rm tomog}-\mathcal E_{\rm in\text{-}situ}$ for the
finite-shot runs, where positive values mean
in-situ training is better. Markers show medians and error bars span the
25th--75th percentiles over 30 targets.}
    \label{fig:insitu_vs_tomography_calibration}
\end{figure}

We address the following questions in numerical simulations. First, does the condition
number predict how much errors are amplified
(Fig.~\ref{fig:nulling_conditioning})?
Second, how well does training work with a finite number of measurement shots
(Tables~\ref{tab:nulling-loss} and~\ref{tab:scaling-N})?
Third, what advantage does in-situ training offer over tomography followed
by offline compilation
(Fig.~\ref{fig:insitu_vs_tomography_calibration})?

\paragraph{Setup.}
Unless stated otherwise, we sample targets in Bloch--Messiah form,
$S_{\rm unk}=O_1K(\mathbf r)O_2$, where $O_1,O_2$ are Haar-random passive
transformations and $K(\mathbf r)$ squeezes mode $i$ by
$r_i\sim\operatorname{Uniform}[0,0.5]$~\cite{bloch1962canonical,fabre2020modes}.
This gives up to $4.3$~dB per mode, which is within experimentally demonstrated
squeezing levels~\cite{clark2026integrated,vahlbruch2016detection}.
We report the normalized residual
$\mathcal E=\|S_\theta S_{\rm unk}-I\|_F/\sqrt{2N}$, which vanishes for
perfect inversion. Finite-shot simulations include noise in homodyne measurements, with each
of the settings in Section~\ref{sec:architecture} averaged over $q$ shots. Appendix~\ref{app:experimental-setup} gives the noise model and simulation details of all experiments.

\paragraph{Conditioning tracks error amplification (Fig.~\ref{fig:nulling_conditioning}).}
To separate the effect of conditioning from optimizer behavior and shot
noise, we use a standard sensitivity test based on small input
perturbations~\cite{higham2002accuracy}. We add a random error of fixed
Frobenius norm $\epsilon=10^{-8}$ to each measured block, compute the
closed-form minimum-squeezing gate from the perturbed blocks, and apply
it to the true response. We run three sweeps, shown in the three columns
of Fig.~\ref{fig:nulling_conditioning}. The first increases the
number of modes $N$. The second follows paths
$S(\alpha)=e^{\alpha L}S_{\rm iso}$, where $S_{\rm iso}$ has an exactly
failing pivot and $L$ is a symplectic generator constructed from a random
Bloch--Messiah target. This construction preserves symplecticity. As $\alpha$ decreases toward
zero, the target approaches $S_{\rm iso}$. The third increases
$s_{\max}$, the upper bound on the target's single-mode Bloch--Messiah
squeezing in dB, at $N=4$.

In all three sweeps, the final residual largely follows
$\kappa_{\max}$. Near a zero pivot, both the residual and $\kappa_{\max}$ grow roughly as
$\alpha^{-1}$, which is steeper than the nulling-step factor
$\sqrt{\kappa}$ discussed above because later gates in the sequence can
further amplify errors left by earlier steps.
Raising $s_{\max}$ to $10$~dB at $N=4$ increases the residual only mildly, whereas increasing $N$ leads to substantially larger errors, consistent with error accumulation and amplification over more nulling steps. Conditioning also sets a hardware requirement through the largest
local minimum-squeezing parameter over the two-mode nulling gates,
$\tfrac12\operatorname{arccosh}\kappa_{\max}$.
For each target, we take the largest squeezing required by any nulling
step. For targets with $r_i\sim\operatorname{Uniform}[0,0.5]$ (up to
$4.3$~dB per Bloch--Messiah mode), the median of this value increases from
$1.1$~dB at $N=2$ to $19.7$~dB at $N=30$ (25th--75th percentiles
$16.7$--$25.0$~dB). This exceeds the squeezing levels available in current
experimental setups~\cite{hirota2026generation,vahlbruch2016detection,karnik2026onchip}.
Although this squeezing requirement becomes challenging at larger $N$, current squeezing-gate
platforms involve only a few
modes~\cite{yoshikawa2008demonstration,larsen2021deterministic}. At $N=4$
with $s_{\max}=10$~dB, the median requirement is about $4.5$~dB, comparable
to demonstrated squeezing gates. The squeezing requirement also grows near nulling
failures. Monitoring $\kappa$
during training can identify these poorly conditioned steps, and passive
mode mixing before nulling can reduce the squeezing required for nulling
(Appendix~\ref{app:universal}).

\begin{table}[t]
\centering
\footnotesize
\setlength{\tabcolsep}{2pt}
\caption{Local loss $\|M_{j,k}\|_F^2$ left by each two-mode gate for
$N=3$, measured on the exact response right after the gate is set. Here
$q$ is the number of shots per homodyne setting, and $q=\infty$ means
noiseless measurements, where the closed-form loss is at machine precision.
All four methods use matched total measurement budgets
(Appendix~\ref{app:finite-shot}). Entries show the median [25th, 75th
percentile] over 90 gates from 30 targets, and each power of ten applies to
all three values.}
\begin{tabular}{lcccc}
\toprule
$q$ & closed-form & finite differences & dithering & genetic algorithm \\
\midrule
$10^{2}$ & $8.2\,[4.5,12]\,10^{-6}$ & $5.4\,[1.3,17]\,10^{-3}$ & $3.2\,[1.1,7.9]\,10^{-3}$ & $6.7\,[3,15]\,10^{-3}$ \\
$10^{3}$ & $7.7\,[4.7,13]\,10^{-7}$ & $2.5\,[0.83,5.3]\,10^{-4}$ & $2.2\,[1.4,3.6]\,10^{-4}$ & $6.5\,[3.3,12]\,10^{-4}$ \\
$10^{4}$ & $6.3\,[4.3,12]\,10^{-8}$ & $5.5\,[3,12]\,10^{-5}$ & $1\,[0.5,1.9]\,10^{-4}$ & $1.6\,[0.92,2.4]\,10^{-4}$ \\
$10^{5}$ & $8.3\,[4.5,12]\,10^{-9}$ & $2.9\,[1.3,7.8]\,10^{-5}$ & $6.5\,[3.2,12]\,10^{-5}$ & $1.2\,[0.49,1.8]\,10^{-4}$ \\
$\infty$ & $7.2\,[3.1,26]\,10^{-31}$ & $2.2\,[1.2,6.6]\,10^{-5}$ & $6.3\,[3.3,15]\,10^{-5}$ & $9\,[5.8,16]\,10^{-5}$ \\
\bottomrule
\end{tabular}
\label{tab:nulling-loss}
\end{table}

\begin{table}[t]
\centering
\footnotesize
\setlength{\tabcolsep}{2pt}
\caption{Final full-mesh residual $\mathcal E$ for increasing $N$, with
$q=10^3$ shots per homodyne setting. All four methods use matched total
measurement budgets (Appendix~\ref{app:finite-shot}). Entries show the
median [25th, 75th percentile] over 30 targets.}
\begin{tabular}{lcccc}
\toprule
$N$ & closed-form & finite differences & dithering & genetic algorithm \\
\midrule
3 & $0.0011\,[0.00099,0.0012]$ & $0.035\,[0.024,0.058]$ & $0.057\,[0.033,0.087]$ & $0.056\,[0.038,0.11]$ \\
5 & $0.0016\,[0.0016,0.0019]$ & $0.098\,[0.069,0.12]$ & $0.086\,[0.07,0.11]$ & $0.11\,[0.073,0.16]$ \\
7 & $0.0022\,[0.0021,0.0025]$ & $0.12\,[0.097,0.15]$ & $0.12\,[0.1,0.18]$ & $0.14\,[0.1,0.2]$ \\
9 & $0.0027\,[0.0023,0.0029]$ & $0.16\,[0.14,0.23]$ & $0.16\,[0.12,0.18]$ & $0.18\,[0.12,0.23]$ \\
11 & $0.0032\,[0.0027,0.0054]$ & $0.2\,[0.17,0.26]$ & $0.16\,[0.14,0.2]$ & $0.18\,[0.15,0.26]$ \\
13 & $0.004\,[0.0031,0.011]$ & $0.22\,[0.19,0.28]$ & $0.22\,[0.16,0.24]$ & $0.21\,[0.17,0.3]$ \\
15 & $0.0046\,[0.0036,0.013]$ & $0.26\,[0.21,0.29]$ & $0.24\,[0.2,0.29]$ & $0.23\,[0.19,0.29]$ \\
20 & $0.013\,[0.0053,0.057]$ & $0.32\,[0.27,0.37]$ & $0.32\,[0.24,0.35]$ & $0.3\,[0.25,0.38]$ \\
25 & $0.021\,[0.0069,0.061]$ & $0.39\,[0.31,0.44]$ & $0.34\,[0.31,0.36]$ & $0.33\,[0.29,0.38]$ \\
30 & $0.047\,[0.02,0.22]$ & $0.47\,[0.41,0.52]$ & $0.41\,[0.36,0.45]$ & $0.36\,[0.31,0.4]$ \\
\bottomrule
\end{tabular}
\label{tab:scaling-N}
\end{table}

\paragraph{Finite-shot training (Tables~\ref{tab:nulling-loss} and~\ref{tab:scaling-N}).}
We compare measured closed-form updates with three feedback methods using matched total measurement budgets. We estimate gradients using finite differences or dithering, which applies small oscillations to the gate parameters and uses demodulation (measuring how the loss responds to each oscillation) to extract gradients, as demonstrated in unitary meshes~\cite{mor2026separating}. Both use Adam~\cite{kingma2017adam} for optimization. The third method is a gradient-free genetic algorithm. All three feedback methods need only the measured loss $\|\widehat M_{j,k}\|_F^2$.
Other gradient methods include automatic differentiation in simulation,
parameter-shift rules~\cite{schuld2019evaluating,jiang2025gradients}, and
in-situ backpropagation~\cite{hughes2018training,pai2023experimentally,ashtiani2026integrated}.

Table~\ref{tab:nulling-loss} compares the four methods on a single nulling
step. For $N=3$, the closed-form squared loss $\|M_{j,k}\|_F^2$ decreases
roughly as $1/q$. The residual norm $\|M_{j,k}\|_F$ instead scales as
$1/\sqrt q$, consistent with the noise-sensitivity bound in
Section~\ref{subsec:conditioning-noise} for fixed conditioning.
Dithering gives lower median local loss than finite differences at
$q=10^2$ and $10^3$, but higher loss at larger $q$.
The feedback optimizers retain a higher loss even without measurement noise
($q=\infty$), so their accuracy is limited by the optimizer or its budget
rather than by shot noise. Table~\ref{tab:scaling-N} reports full-mesh
training for $N=3$ to $30$. The residual grows with $N$ for all four
methods. The median closed-form residual remains below $10^{-2}$ up to
$N=15$, then reaches $0.013$ at $N=20$ and $0.047$ at $N=30$. With
noiseless measurements, the median stays between $1.6\times10^{-15}$ and
$1.4\times10^{-14}$ for all $N$. The growth with $N$ is consistent with the growth of $\kappa_{\max}$ in
Fig.~\ref{fig:nulling_conditioning}. In both tables, total shot budgets are matched to compare the methods.
Because the closed-form method measures each gate once rather than iterating,
it uses about $10^3$ times more shots per setting than each feedback loss
evaluation (Appendix~\ref{app:finite-shot}). In practice, the closed-form method
may need far fewer shots to reach a useful residual. A practical approach may combine the methods,
using the closed-form gate as a starting point and refining it with measured
feedback to correct errors from an imperfect calibration model.

\paragraph{Fixed calibration errors (Fig.~\ref{fig:insitu_vs_tomography_calibration}).}
We now test when in-situ training can outperform tomography followed by
offline compilation. Offline compilation computes all gate settings from a
hardware model and cannot account for unmodeled deviations from those settings.
In-situ training instead measures the response after the real, imperfect gates
have acted. To represent static errors such as fabrication variations and phase
offsets, we add a small symplectic error to each gate, with each error
parameter drawn from
$\operatorname{Uniform}[-\epsilon_{\max},\epsilon_{\max}]$ and fixed during
each trial (Appendix~\ref{app:calibration-error}). The tomography baseline
measures the full response once, projects it onto a symplectic
matrix, and computes all gates offline. In-situ training computes each gate
from blocks measured after the earlier gates are set, using sequential access
(Section~\ref{sec:setup}). Both use the closed-form minimum-squeezing gate and the same measurement budget. With
noiseless measurements (dotted curves in
Fig.~\ref{fig:insitu_vs_tomography_calibration}), in-situ training gives
about half the residual of tomography at every $\epsilon_{\max}$ we test.
We also add a moderate level of shot noise ($q=10^4$) to test when
gate errors outweigh measurement noise.

For small gate errors, measurement noise dominates and tomography is
slightly more accurate. In the left panel of
Fig.~\ref{fig:insitu_vs_tomography_calibration}, the median residuals at
$\epsilon_{\max}=10^{-3}$ are $0.020$ for tomography and $0.027$ for in-situ
training. This advantage comes mainly from the global symplectic projection
of the tomography estimate. Without it, the median tomography residual rises
to $0.026$. In-situ training has no global projection, but its gates are
symplectic by construction. As $\epsilon_{\max}$ grows, gate errors
overtake shot noise. The two methods cross over at
$\epsilon_{\max}\approx4\times10^{-3}$, and from
$\epsilon_{\max}\approx2\times10^{-2}$ onward the in-situ median residual is
about half that of tomography, matching the noiseless trend. The right panel shows
$\mathcal E_{\rm tomog}-\mathcal E_{\rm in\text{-}situ}$, the tomography
residual minus the in-situ residual for each target. It stays close to zero
for small gate errors and grows steadily once gate errors dominate.

\section{Discussion and limitations}
\label{sec:discussion}

\paragraph{Limitations.}
A key obstacle to a hardware
demonstration is likely to be loss, since the theory assumes lossless
operations, whereas loss remains a major challenge for integrated
continuous-variable platforms~\cite{clark2026integrated}. Uniform loss
across modes does not prevent the trained mesh from separating the modes,
but unequal loss leaves residual mixing. This mixing can be removed by
equalizing the loss with dummy components~\cite{miller2015perfect}, at the cost of lower
transmission (Appendix~\ref{app:loss}). Loss also removes quantum
information and adds noise that no mesh can
undo~\cite{weedbrook2012gaussian,serafini2017quantum}. Another difficulty is
that, although nulling fails only for a measure-zero set of targets, targets
near this set have near-zero pivots, which require strong squeezing and
amplify measurement errors. Errors and the required squeezing also grow with
$N$, which may limit scaling to larger meshes (Section~\ref{sec:experiments}). Adding extra passive gates before each nulling step can reduce the required
squeezing, but it costs more components
(Appendix~\ref{app:universal}). If the response can only be measured at the output of the full mesh rather than directly after each gate, every measurement passes through the gates that have not yet been trained, which may also increase errors. Hardware error-correction methods developed
for unitary
meshes~\cite{miller2015perfect,bandyopadhyay2021hardware,hamerly2022accurate,hamerly2022stability}
may also help with these challenges, and adapting them to symplectic meshes
is an open direction. 

\paragraph{Architecture variants.}
Other nulling meshes are possible. For example, a whole column could be
nulled at once rather than one block at a time. Adding extra passive gates before each nulling step, similar to row
pivoting in Gaussian elimination~\cite{golub2013matrix}, can make the mesh
universal~\cite{singer2003rounding} (Appendix~\ref{app:universal}). Other architectures used for unitary and
general linear meshes may also be worth exploring in the symplectic setting,
such as rectangular layouts~\cite{clements2016optimal}, SVD-style
architectures~\cite{miller2013self}, and more.

\paragraph{Beyond in-situ inversion.}
Self-configuring unitary meshes have been used for more than inversion.
They can implement arbitrary linear transformations~\cite{miller2013self}
and perform variational diagonalization of scrambled
light~\cite{roquescarmes2024measuring,roquescarmes2025automated,mor2026separating,karnieli2025variational}.
Symplectic versions of these tasks, which would also allow squeezing,
are a natural next step. More broadly, machine learning often uses structured factorizations and
structure-preserving parameterizations of large linear
transformations~\cite{dao2019learning,arjovsky2016unitary,wisdom2016full,jing2017tunable,mhammedi2017efficient,lezcanocasado2019cheap,frerix2019approximating,shalit2014coordinate,han2024riemannian}.
Unitary Reck meshes provide a photonic realization of this idea and are
used as linear layers in optical neural
networks~\cite{shen2017deep,harris2018linear}. A symplectic Reck mesh
may play a similar role in Gaussian quantum optics and other problems
involving symplectic transformations~\cite{weedbrook2012gaussian,houde2024matrix,cariolaro2018implementation,killoran2019continuous}.

\section{Conclusion}
\label{sec:conclusion}
We introduced the symplectic Reck mesh for learning the inverse of an
unknown lossless Gaussian operation through local optical measurements.
The mesh extends the Reck architecture to include squeezing.
We proved that sequential block nulling in the adjacent-mode Reck ordering
succeeds for generic symplectic matrices and identified exactly when a
nulling step fails. We showed that the condition number $\kappa$ determines the
minimum squeezing required for nulling and bounds the amplification
of measurement errors in one nulling step. Our simulations demonstrate
finite-shot training and show that in-situ updates can partly
compensate for fixed gate errors, giving lower residuals than
tomography followed by offline compilation when these gate errors are
sufficiently large. Potential applications include compensating unwanted mode mixing and
squeezing in continuous-variable optical
links~\cite{annoni2017unscrambling,filip2005elimination,kovalenko2021cross,usenko2026continuous}.
More broadly, the architecture and training methods may enable new
self-configuring tasks in continuous-variable quantum optics.

\begin{ack}
This work is supported by the Toyota Research Institute of North America. The authors are seeking patent protection for inventions related to this
work (US Provisional Patent Application 64/075,643, Stanford University).
\end{ack}

\clearpage
\appendix
\etocdepthtag.toc{appendix}
\etocsettagdepth{main}{none}
\etocsettagdepth{appendix}{section}
\etocsettocstyle{\section*{Appendix contents}}{}
{\small\tableofcontents}

\section{Explicit gate parameterizations, and the $N$-mode mesh}
\label{SM:operations}

\subsection{Explicit gate parameterizations}
\label{app:gates}
We use the mode-grouped quadrature ordering of Section~\ref{sec:setup}, so a
two-mode gate on modes \(a\) and \(b\) acts on \((x_a,p_a,x_b,p_b)^\top\).
Every gate has independently adjustable parameters. We write \(S^{(a,b)}\)
for a generic two-mode gate on modes \(a\) and \(b\), and \(S^{(m)}\) for a
generic single-mode gate on mode \(m\).

Following Ref.~\cite{pai2019matrix}, an arbitrary two-mode unitary can be
written as
\begin{equation}
U(\theta,\phi,\alpha,\beta)
= D(\phi,0)B D(\theta,0)B D(\alpha,\beta)
= i e^{i\theta/2}
\begin{pmatrix}
e^{i(\phi+\alpha)}\sin\tfrac{\theta}{2}
  & e^{i(\phi+\beta)}\cos\tfrac{\theta}{2} \\
e^{i\alpha}\cos\tfrac{\theta}{2}
  & -e^{i\beta}\sin\tfrac{\theta}{2}
\end{pmatrix},
\label{eq:U2full}
\end{equation}
where \(D(\phi,\psi)=\operatorname{diag}(e^{i\phi},e^{i\psi})\) and
\(B=\tfrac{1}{\sqrt2}\left(\begin{smallmatrix}1&i\\i&1\end{smallmatrix}\right)\).
A unitary \(U=(u_{mn})\) corresponds in the real quadrature basis to the
orthosymplectic matrix
\begin{equation}
O(U)=
\begin{pmatrix}
\operatorname{Re}u_{11}&-\operatorname{Im}u_{11}&\operatorname{Re}u_{12}&-\operatorname{Im}u_{12}\\
\operatorname{Im}u_{11}& \operatorname{Re}u_{11}&\operatorname{Im}u_{12}& \operatorname{Re}u_{12}\\
\operatorname{Re}u_{21}&-\operatorname{Im}u_{21}&\operatorname{Re}u_{22}&-\operatorname{Im}u_{22}\\
\operatorname{Im}u_{21}& \operatorname{Re}u_{21}&\operatorname{Im}u_{22}& \operatorname{Re}u_{22}
\end{pmatrix}.
\label{eq:orthosymp}
\end{equation}

\begin{figure}[H]
\centering
\includegraphics[width=0.6\linewidth]{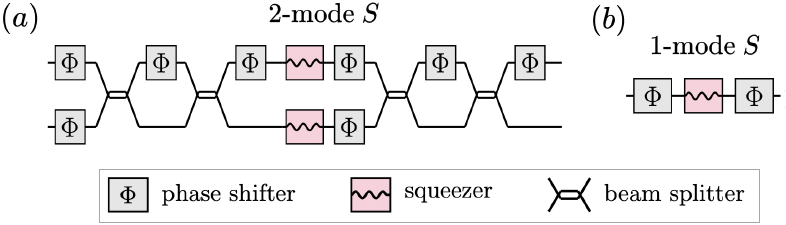}
\caption{Bloch--Messiah implementations of (a) a two-mode symplectic gate
and (b) a single-mode symplectic gate, using phase shifters, squeezers,
and beam splitters.}
\label{fig:gate-bm}
\end{figure}

A general lossless two-mode symplectic gate admits the Bloch--Messiah form
\begin{equation}
S^{(a,b)}=O^{(a,b)}_1
\operatorname{diag}(e^{r_1},e^{-r_1},e^{r_2},e^{-r_2})
O^{(a,b)}_2,
\label{eq:two-mode-bm}
\end{equation}
where the passive factors are implemented with beam splitters and phase
shifters~\cite{bloch1962canonical,fabre2020modes,cariolaro2022implementation}.
The two passive factors and two squeezers give ten real parameters, matching
\(\dim\operatorname{Sp}(4,\mathbb R)=10\). Similarly,
\begin{equation}
S^{(m)}=\mathcal R(\varphi_1)
\begin{pmatrix}e^r&0\\0&e^{-r}\end{pmatrix}
\mathcal R(\varphi_2),
\qquad
\mathcal R(\varphi)=
\begin{pmatrix}\cos\varphi&-\sin\varphi\\
\sin\varphi&\cos\varphi\end{pmatrix},
\end{equation}
which gives the three parameters of \(\operatorname{Sp}(2,\mathbb R)\). Figure~\ref{fig:gate-bm}
illustrates these Bloch--Messiah implementations.

\subsection{Symplectic Reck mesh for \texorpdfstring{$N$}{N} modes}\label{app:general-N}
Using these single-mode and two-mode gates, we now write the full mesh
explicitly as a product of its layers. Layer
$L_k$, for $k=1,\ldots,N-1$, contains the $N-k$ two-mode gates that null
column $k$ (Section~\ref{sec:architecture}), each a gate $S^{(a,a+1)}$ on
adjacent modes, and the final layer $L_N$ applies one single-mode gate
$S^{(m)}$ to each mode $m$, with both gate types defined in
Appendix~\ref{app:gates}. With the rightmost factor acting first,
\begin{equation}
    L_k = S_{L_k}^{(k,k+1)}S_{L_k}^{(k+1,k+2)}\cdots S_{L_k}^{(N-1,N)},
    \qquad
    L_N=\bigoplus_{m=1}^{N}S_{L_N}^{(m)}.
\end{equation}
The full mesh is then
\begin{equation}
    S_\theta=L_NL_{N-1}\cdots L_1 .
\end{equation}

\section{Proof of the symplectic Reck factorization}
\label{app:genericity}

This appendix proves the local nulling criterion
(Section~\ref{subsec:local-pivot}) and Theorem~\ref{thm:generic}. We first
establish when one step succeeds, then show that the full sequence
completes the inversion for generic targets.

\subsection{One nulling step}
\label{app:genericitygate}
With $P$, $E$ and $\Delta$ as in Section~\ref{subsec:local-pivot}, let
$V=\left(\begin{smallmatrix}P\\E\end{smallmatrix}\right)$. We seek a
symplectic gate $G$ with
$GV=\left(\begin{smallmatrix}Q\\0\end{smallmatrix}\right)$.
For any real $2\times2$ matrix $X$,
\begin{equation}
X^\top J_2X=(\det X)\,J_2.
\label{eq:area-identity}
\end{equation}
Applying this to $P$ and $E$ gives
\begin{equation}
V^\top J_4V=P^\top J_2P+E^\top J_2E=\Delta J_2,
\end{equation}
which is preserved by any $G$ with $G^\top J_4G=J_4$.

If $V=0$, the target block is already zero. If $V$ has rank one, both
columns are multiples of the same real four-vector. Pairing its $x$ and
$p$ components gives one complex two-mode amplitude, which a beam
splitter with phase shifters moves into the upper mode.

Now suppose $V$ has rank two. If
$GV=\left(\begin{smallmatrix}Q\\0\end{smallmatrix}\right)$, then
$Q^\top J_2Q=\Delta J_2$, so $\det Q=\Delta$. Since $G$ is invertible,
$Q$ has rank two, so nulling requires $\Delta\neq0$.

Conversely, suppose $\Delta\neq0$ and that $P$ is invertible. The gate
$\left(\begin{smallmatrix}I_2&0\\C&I_2\end{smallmatrix}\right)$ with
$C=-EP^{-1}$ nulls $E$ but is not in general symplectic. We therefore
add a free block $B$,
\begin{equation}
G_0=\begin{pmatrix}I_2&B\\C&I_2\end{pmatrix},
\qquad
G_0V=\begin{pmatrix}P+BE\\0\end{pmatrix},
\end{equation}
which still nulls $E$ for any $B$, and choose $B$ to make $G_0$
symplectic. We have
\begin{equation}
G_0^\top J_4G_0=
\begin{pmatrix}
J_2+C^\top J_2C&J_2B+C^\top J_2\\
B^\top J_2+J_2C&B^\top J_2B+J_2
\end{pmatrix}.
\end{equation}
Choosing $B=J_2C^\top J_2$ makes the off-diagonal blocks vanish. Using
Eq.~\eqref{eq:area-identity} and $\det B=\det C$ then gives
\begin{equation}
G_0^\top J_4G_0=\lambda J_4,
\qquad
\lambda=1+\det C=1+\frac{\det E}{\det P}
=\frac{\Delta}{\det P}\neq0.
\end{equation}
To remove this scalar factor, set $L=\operatorname{diag}(1,\lambda^{-1})$,
so $L^\top J_2L=\lambda^{-1}J_2$, and define
\begin{equation}
G=(L\oplus L)G_0.
\label{eq:shear-gate}
\end{equation}
Then $G^\top J_4G=J_4$ and
$GV=\left(\begin{smallmatrix}L(P+BE)\\0\end{smallmatrix}\right)$, so $G$
is symplectic and nulls $E$. If $P$ is singular, then $E$ is invertible
since $\det E=\Delta\neq0$. A passive swap of the two modes, absorbed
into the same gate, reduces this case to the one above.

Hence, for rank-two $V$, a symplectic nulling gate exists if and only if
$\Delta\neq0$, and for rank at most one it always exists. Similar criteria
were derived in Refs.~\cite{singer2012orthosymmetric,langer2026matchgate}
using different explicit gates, assuming invertible $P$ or invertible $P$
and $E$, respectively.

\subsection{Preservation of zeros and block diagonality}
\label{app:preservation}
Having shown that each step can null its target block, we now show that
later steps do not undo earlier ones, and that the result is block
diagonal. At nulling step $(k,j)$ the gate mixes only block rows $j-1$ and $j$. In
column $k$, the blocks below row $j$ are already zero and are not touched
again. In every earlier column, the blocks in rows $j-1$ and $j$ are already
zero, so mixing these rows preserves both zeros. After all $N(N-1)/2$
two-mode steps, the response is block upper triangular. A block upper-triangular symplectic matrix is block diagonal. Write
$T=\left(\begin{smallmatrix}T_1&R\\0&T_{\rm rem}\end{smallmatrix}\right)$.
The top-left and top-right parts of $T^\top J_{2N}T=J_{2N}$ give
$T_1^\top J_2T_1=J_2$ and $T_1^\top J_2R=0$. The first makes $T_1$
invertible, so the second gives $R=0$. Repeating this for $T_{\rm rem}$ gives
$T=T_1\oplus\cdots\oplus T_N$ with each $T_m\in\operatorname{Sp}(2,\mathbb R)$,
and the final layer applies $T_m^{-1}$ on each mode, giving
$S_\theta S_{\mathrm{unk}}=I$.

\subsection{The choice of nulling gate does not change later pivots}
\label{app:pivot-independence}
The gate that completes a nulling step is not unique. We now show that
the choice of nulling gate does not affect whether later steps succeed, a
result we use in the genericity proof below. Suppose $G$ and $\widetilde G$ both null the same rank-two response $V$,
giving invertible upper blocks $Q$ and $\widetilde Q$. Then
$H=G\widetilde G^{-1}$ satisfies
\begin{equation}
H\begin{pmatrix}\widetilde Q\\0\end{pmatrix}
=\begin{pmatrix}Q\\0\end{pmatrix}.
\end{equation}
Since $\widetilde Q$ is invertible, $H$ maps the upper mode into itself, so
$H$ is block upper triangular. By Appendix~\ref{app:preservation},
$H=A\oplus D$ with $A,D\in\operatorname{Sp}(2,\mathbb R)$, so two valid
nulling gates differ only by single-mode gates on their outputs. Thus, changing
the gate replaces each block $X$ it acts on with $AX$ or $DX$.
Since $\det A=\det D=1$, every block determinant (and hence every later
pivot) is unchanged. This means that whether the full sequence succeeds depends
only on the entries of $S_{\rm unk}$. The choice of gate does still affect the condition number $\kappa$ of later
steps, because single-mode squeezing changes the Euclidean area spanned by
the response columns.

\subsection{Generic success: proof of Theorem~\ref{thm:generic}}
\label{app:generic-proof}

The previous subsections show that the mesh inverts a target whenever every nulling
step has $\Delta_t\neq0$. We now show that this holds for
almost every target. We do this by showing that
every failing target lies on the zero set of a single nonzero polynomial. By Appendix~\ref{app:pivot-independence}, we may use the gate
of Eq.~\eqref{eq:shear-gate} at every step. We call a target
\emph{exceptional} if $\det P_t=0$ or $\Delta_t=0$ at some step $t$. This
set contains every failure. It also contains some targets that can still be
nulled (those with $\det P_t=0$ but $\Delta_t\neq0$). This does not matter since including
these only enlarges the exceptional set. Showing that this larger set has measure zero still shows that the failures have measure zero. Outside this set, each gate is built from
the current response using only additions, multiplications, determinants
and $P_t^{-1}$. Every later entry, and every $\det P_t$ and
$\Delta_t$, is therefore a ratio of polynomials in the entries of
$S_{\rm unk}$. Let $\Pi$ be the product of all these numerators and
denominators. Every exceptional target satisfies $\Pi(S_{\rm unk})=0$.

It remains to show that $\Pi$ is not identically zero. One target with
$\Pi\neq0$ is enough to show this. To build such an example, let us start from a diagonal unitary and apply
$50{:}50$ beam splitters in the reverse order of the nulling steps. The
nulling sequence then removes these beam splitters one at a time. Let $p_t$ and $e_t$
be the complex entries corresponding to $P_t$ and $E_t$. Each $50{:}50$
beam splitter maps a nonzero entry to two nonzero entries of equal
magnitude, so $p_t\neq0$ at every step. For a passive matrix, a complex
entry $z$ becomes a $2\times2$ block with determinant $|z|^2$, so
\begin{equation}
\det P_t=|p_t|^2>0,\qquad \Delta_t=|p_t|^2+|e_t|^2>0.
\end{equation}
By Appendix~\ref{app:pivot-independence}, the gate of
Eq.~\eqref{eq:shear-gate} encounters the same block determinants, so
$\Pi\neq0$ for this target. The zero set of $\Pi$, which contains every
exceptional target, therefore has measure zero in
$\operatorname{Sp}(2N,\mathbb R)$.

Finally, the same result gives a factorization theorem. Applying the inversion
procedure to $S^{-1}$ gives a mesh with $S_\theta S^{-1}=I$, so
$S_\theta=S$. Since inversion is a diffeomorphism of
$\operatorname{Sp}(2N,\mathbb R)$, it preserves genericity, so generic
symplectic matrices admit a symplectic Reck factorization.

\section{Universal architecture}
\label{app:universal}
Theorem~\ref{thm:generic} shows that the symplectic Reck mesh inverts
generic targets. Similar to Ref.~\cite{singer2003rounding}, we add passive
gates before each two-mode symplectic nulling step, which avoids zero
pivots. Here we give one example of a modified version of the symplectic
Reck mesh that inverts every $S\in\operatorname{Sp}(2N,\mathbb R)$. Before nulling $M_{j,k}$ with
$j>k+1$, we add extra passive two-mode gates on modes
$(k,k+1),(k+1,k+2),\ldots,(j-2,j-1)$, applied in that order, and then the
usual symplectic gate acts on modes $(j-1,j)$. Each extra gate is a tunable
beam splitter with a phase shifter, a special case of the two-mode unitary
of Eq.~\eqref{eq:U2full}. The gate on modes $(a,a+1)$ is set so that mode
$a$ no longer responds to the $x_k$ probe, moving that response into mode
$a+1$, exactly as a gate of the unitary Reck mesh moves light from one
waveguide into the next. No extra gates are needed when $j=k+1$. As before,
single-mode symplectic gates complete the inversion.

To see why this works, let $M$ be the current response before nulling
$M_{j,k}$. The modes before $k$ are already decoupled
(Appendix~\ref{app:preservation}), and the blocks below row $j$ have been
nulled, so only rows $k,\ldots,j$ of column $k$ are nonzero. Since $M$ is
symplectic, the diagonal block of $M^\top J_{2N}M=J_{2N}$ for column $k$,
together with Eq.~\eqref{eq:area-identity}, gives
\begin{equation}
\sum_{i=k}^{j}\det M_{i,k}=1.
\end{equation}
Intuitively, this equation says that input mode $k$ carries one unit of
symplectic area (Section~\ref{subsec:conditioning-noise}), spread over
output modes $k,\ldots,j$. The extra gates collect this area into modes
$j-1$ and $j$. After them, the blocks $M_{i,k}$ with $k\leq i\leq j-2$ have
a zero first column and hence zero determinant. The extra gates are
symplectic and leave mode $j$ unchanged, so
\begin{equation}
\Delta=\det M_{j-1,k}+\det M_{j,k}=1,
\end{equation}
and the gate on modes $(j-1,j)$ can null $M_{j,k}$
(Section~\ref{subsec:local-pivot}). For $j=k+1$, the same identity gives
$\Delta=1$ without extra gates. The extra gates act only on modes
$k,\ldots,j-1$, whose blocks in earlier columns are already zero, so they
preserve every earlier zero. Appendix~\ref{app:preservation} then completes
the inversion.

In hardware, each extra gate can be set by feedback, minimizing the measured
response of its first mode to the $x_k$ probe, or computed in closed form
from that response. In noiseless simulations with the minimum-squeezing
gate, the extra gates also reduce the squeezing required by the two-mode
gates, $\tfrac12\operatorname{arccosh}\kappa_{\max}$. For 30 random targets with $N=20$, the median $\kappa_{\max}$ drops from
$7.4$ to $1.04$, corresponding to $11.7$~dB and $1.3$~dB of squeezing. As the target approaches a nulling failure in Fig.~\ref{fig:nulling_conditioning}, the required squeezing remains near $5$~dB with the extra gates, whereas without them it grows to tens of dB. Other constructions of universal
architectures are also possible.

\section{Conditioning and minimum squeezing}
\label{app:minimal-gate}

The nulling criterion establishes when a gate exists. We now ask how much
squeezing it requires and how strongly it amplifies measurement errors.
We construct a minimum-squeezing gate, prove its optimality, count its
parameters, and derive the nulling-step error bound of
Section~\ref{subsec:conditioning-noise}. We assume
$V$ has rank two and $\Delta\neq0$, with $\kappa$ as in Eq.~\eqref{eq:kappa}.

\subsection{The minimum-squeezing gate}
We build the gate from a two-mode passive gate $R$ followed by the two-mode
squeezer
\begin{equation}
T_{\rm sq}(r)=\begin{pmatrix}\cosh r\,I_2&\sinh r\,Z\\ \sinh r\,Z&\cosh r\,I_2\end{pmatrix},
\qquad Z=\operatorname{diag}(1,-1).
\end{equation}
This squeezer can itself be built from two single-mode squeezers between two
passive interferometers~\cite{weedbrook2012gaussian}. For $T_{\rm sq}(r)$ to
output $\left(\begin{smallmatrix}Q\\0\end{smallmatrix}\right)$, its input must be
\begin{equation}
W=T_{\rm sq}(r)^{-1}\begin{pmatrix}Q\\0\end{pmatrix}
=\begin{pmatrix}\cosh r\,Q\\-\sinh r\,ZQ\end{pmatrix},
\end{equation}
so the passive gate must map $V$ to $W$.

We first choose $r$ and $Q$. A passive gate preserves both $V^\top V$ and
$V^\top J_4V=\Delta J_2$, so $W$ must have the same two products. Using
$Z^\top J_2Z=-J_2$,
\begin{equation}
W^\top W=\cosh(2r)\,Q^\top Q,
\qquad
W^\top J_4W=(\det Q)\,J_2.
\end{equation}
We therefore need $Q^\top Q=V^\top V/\cosh(2r)$ and $\det Q=\Delta$.
Taking determinants gives $\cosh(2r)=\kappa$, so
\begin{equation}
r=\tfrac12\operatorname{arccosh}\kappa.
\label{eq:r-min}
\end{equation}
Let $Q_0=(V^\top V/\kappa)^{1/2}$ be the symmetric positive square root,
so $Q_0^\top Q_0=V^\top V/\kappa$ and $\det Q_0=|\Delta|$. We take $Q=Q_0$
if $\Delta>0$ and $Q=ZQ_0$ if $\Delta<0$, which leaves $Q^\top Q$
unchanged and gives $\det Q=\Delta$.

This requires $\kappa\geq1$, which always holds. To see this, define the
complex $2\times2$ matrix $A_V$ with entries
$(A_V)_{mc}=V_{x_m,c}+iV_{p_m,c}$, so that each row gives the complex
amplitude $x_m+ip_m$ of output mode $m$ in response to input quadrature
$c$. A direct calculation gives
\begin{equation}
A_V^\dagger A_V=V^\top V+i\,V^\top J_4V=V^\top V+i\Delta J_2.
\label{eq:complex-gram}
\end{equation}
Since $A_V^\dagger A_V$ is positive semidefinite, its determinant
$\det(V^\top V)-\Delta^2$ is nonnegative, so $\kappa\geq1$.

With $r$ and $Q$ fixed, it remains to find a passive gate $R$ that maps
$V$ to $W$. Define $A_W$ from $W$ in the same way. Since $W$
has the same two products as $V$, Eq.~\eqref{eq:complex-gram} gives
$A_W^\dagger A_W=A_V^\dagger A_V$. If $\kappa>1$, then
$|\det A_V|^2=\det(V^\top V)-\Delta^2>0$, so $A_V$ is invertible and
$U=A_WA_V^{-1}\in\mathrm{U}(2)$ satisfies $UA_V=A_W$. We take $R=O(U)$,
the two-mode passive gate that applies $U$ to the complex amplitudes
$x_m+ip_m$ of the two modes (Eq.~\eqref{eq:orthosymp}). Any element of
$\mathrm{U}(2)$ can be realized with a beam splitter and phase shifters, as
in Eq.~\eqref{eq:U2full} (Appendix~\ref{SM:operations}). Then $R$ maps
$A_V$ to $UA_V=A_W$, so $RV=W$. If $\kappa=1$, then $r=0$ and the columns
of $A_V$ are proportional, so a beam splitter with phase shifters moves
both into the upper mode. In either case, $G=T_{\rm sq}(r)R$ nulls $E$.

\subsection{No nulling gate uses less squeezing}
The construction above achieves squeezing $r$. To show that this is
minimal, we compare it with an arbitrary nulling gate. By Appendix~\ref{app:pivot-independence}, any other nulling gate has the form $\widetilde G=(L_1\oplus L_2)T_{\rm sq}(r)R$ with $L_1,L_2\in\operatorname{Sp}(2,\mathbb R)$. Since $R$ is orthogonal,
\begin{equation}
\|\widetilde G\|_F^2=\cosh(2r)\bigl(\|L_1\|_F^2+\|L_2\|_F^2\bigr)\geq4\cosh(2r),
\end{equation}
because each $L_i$ has singular values $\sigma,\sigma^{-1}$, so
$\|L_i\|_F^2=\sigma^2+\sigma^{-2}\geq2$. If $\widetilde G$ has
Bloch--Messiah squeezing parameters $s_1,s_2\geq0$
(Eq.~\eqref{eq:two-mode-bm}) and $s=\max(s_1,s_2)$, then
\begin{equation}
\|\widetilde G\|_F^2=2\cosh 2s_1+2\cosh 2s_2\leq4\cosh 2s.
\end{equation}
Together these give $s\geq r$. Our gate squeezes by exactly $r$, so it
minimizes the largest squeezing parameter.

\subsection{Four parameters per gate}
\label{app:four-parameters}
We next count the independent controls needed to implement this gate.
Naively, each minimum-squeezing gate has five parameters, one squeezing
parameter $r$ and four in the passive part $R$, but one is redundant. Let
$D_\phi=\mathcal R(\phi)\oplus\mathcal R(-\phi)$. Since
$\mathcal R(\phi)Z=Z\mathcal R(-\phi)$, $D_\phi$ commutes with
$T_{\rm sq}(r)$, so
\begin{equation}
T_{\rm sq}(r)D_\phi RV=D_\phi\begin{pmatrix}Q\\0\end{pmatrix}=\begin{pmatrix}\mathcal R(\phi)Q\\0\end{pmatrix}.
\end{equation}
Replacing $R$ by $D_\phi R$ therefore still nulls $E$, so this phase can be
fixed. With four parameters per two-mode gate and three per single-mode gate,
\begin{equation}
4\cdot\frac{N(N-1)}{2}+3N=N(2N+1)=\dim\operatorname{Sp}(2N,\mathbb R),
\end{equation}
matching the dimension of the symplectic group~\cite{arvind1995real}.

\subsection{Nulling-step error amplification}
\label{app:error-amplification}
The same gate construction also gives a direct bound on the residual
left by measurement errors. Suppose the gate $\widehat G$ is computed from a noisy estimate
$\widehat V=V+\delta V$ with nonzero estimated pivot, and let $\widehat N$ be
the bottom two rows of $\widehat G$. By construction
$\widehat N\widehat V=0$, so on the true response
\begin{equation}
\widehat NV=\widehat N(\widehat V-\delta V)=-\widehat N\,\delta V.
\end{equation}
The bottom two rows of $T_{\rm sq}(\widehat r)$ are orthogonal with squared
norm $\cosh^2\widehat r+\sinh^2\widehat r=\cosh 2\widehat r=\widehat\kappa$,
and right multiplication by the orthogonal $\widehat R$ preserves their inner
products. Hence $\widehat N\widehat N^\top=\widehat\kappa I_2$, and
\begin{equation}
\|\widehat NV\|_F\leq\sqrt{\widehat\kappa}\,\|\delta V\|_F,
\label{eq:local-error-amplification}
\end{equation}
with equality in the worst case, since both singular values of $\widehat N$
equal $\sqrt{\widehat\kappa}$.

\section{Training cost}
\label{app:training-cost}
We now count the measurements and computation required to train the full
mesh. In our experiments, we choose a fixed measurement budget for each nulling
step, with $q$ shots per homodyne setting and, for feedback training,
$n_{\rm eval}$ loss evaluations per gate (Appendix~\ref{app:finite-shot}).
Another option could have been to train each gate until its measured loss
falls below a chosen threshold, spending fewer evaluations on easy steps.
The mesh has $N(N-1)/2$ two-mode gates and $N$ single-mode gates, each with
a constant number of parameters, so it has $O(N^2)$ parameters. Closed-form
training uses $16$ homodyne settings per two-mode gate and $8$ per
single-mode gate, or $16\cdot N(N-1)/2+8N=8N^2$ settings in total, which is
$8qN^2$ shots. Feedback training uses $8$ settings per loss evaluation, so
it uses $O(n_{\rm eval}qN^2)$ shots. Each closed-form update is a
calculation on two $2\times2$ blocks, so the classical computation is
$O(N^2)$, and the training time also scales as $O(N^2)$. For comparison, tomography followed by offline compilation also needs
$O(qN^2)$ shots when blocks are measured one at a time, which is the matched
budget of Fig.~\ref{fig:insitu_vs_tomography_calibration}. If all outputs
are measured in parallel, tomography recovers a whole block column per
probe, reducing this to $O(qN)$, whereas in-situ training remains $O(qN^2)$
because each update depends on the previous one. Tomography then needs
$O(N^3)$ classical computation to invert and decompose the estimate. In
practice, training time is likely to be limited by measurement and hardware
reconfiguration rather than by classical computation.

\subsection{Conditional sample-complexity guarantee}
\label{app:pac}
The cost above is expressed in terms of the shots per setting, $q$. We
now bound how large $q$ must be for closed-form training to reach
a residual $\mathcal E\leq\epsilon$ with probability at least $1-\delta$,
by combining the nulling-step bound of Section~\ref{subsec:conditioning-noise}
over the full nulling sequence. We assume exactly implemented gates, so
the only errors come from finite-shot measurements, together with a set
of stability conditions stated below. Let $M=S_\theta S_{\rm unk}$ be the
final response and write $M-I=L+U+D$, where $L$, $U$ and $D$ contain the
blocks below, above and on the block diagonal. Let $R_t$ be the residual
block left immediately after nulling step $t$, and let $H_k$ be the
combined action of all later gates on modes $k+1,\ldots,N$.

We first show how these residuals accumulate. Once column $k$ is
processed, its lower part is a stack of that column's residual blocks,
which later gates multiply by $H_k$. If $\|H_k\|_2\leq\Gamma$ for every
$k$, then $\|L\|_F^2\leq\Gamma^2\sum_t\|R_t\|_F^2$. Symplecticity gives
$M^{-1}-I=-J(M-I)^\top J$, so the upper part of $M^{-1}-I$ has norm
$\|L\|_F$ and $\|M^{-1}-I\|_F=\|M-I\|_F$. Taking the upper part of the
identity $(M-I)+(M^{-1}-I)=-(M-I)(M^{-1}-I)$ gives
$\|U\|_F\leq\|L\|_F+\rho\|M-I\|_F$, where $\rho=\|M-I\|_2$. For $\rho<1$,
\begin{equation}
\mathcal E^2\leq
\frac{4\Gamma^2\sum_t\|R_t\|_F^2+\|D\|_F^2}{N(1-\rho)^2}.
\label{eq:pac-accumulation}
\end{equation}

To turn this accumulation bound into a shot requirement, we impose the
following stability conditions. Suppose that, with probability at least
$1-\delta/2$, every measured
condition number satisfies $\widehat\kappa_t\leq K$, every
$\|H_k\|_2\leq\Gamma$ and $\|M-I\|_2\leq\rho_0<1$, where these constants
are fixed in advance. These conditions are most natural for few-mode meshes, since the worst
condition number and the amplification by later gates tend to grow with $N$
for random targets (Fig.~\ref{fig:nulling_conditioning}), and
$\|M-I\|_2\leq\rho_0<1$ restricts the guarantee to the regime near
convergence. Suppose also that,
on these runs, the mean of each $\|R_t\|_F^2$ and of $\|D\|_F^2/(2N)$ is at
most $c/q$. Summing
Eq.~\eqref{eq:pac-accumulation} over the $N(N-1)/2$ nulling steps and
applying Markov's inequality gives the sufficient condition
\begin{equation}
q\geq\frac{4c\,[\Gamma^2(N-1)+1]}{(1-\rho_0)^2\epsilon^2\delta}
\quad\Longrightarrow\quad
\Pr(\mathcal E\leq\epsilon)\geq1-\delta.
\end{equation}
To relate $c$ to the measurement noise, we use the nulling-step bound of
Eq.~\eqref{eq:local-error-amplification},
$\|R_t\|_F^2\leq\widehat\kappa_t\|\delta V_t\|_F^2$. If each measured
quadrature has variance at most $a/2$, each of the eight entries of
$\widehat V_t$ has variance at most $a/(4\eta^2q)$, so one may take
$c=\max\{4aK/\eta^2,c_{\rm diag}\}$, where $c_{\rm diag}/q$ bounds the
diagonal error. With $8N^2$ settings, the total number of shots is
$O\bigl(c\,(\Gamma^2N^3+N^2)/[(1-\rho_0)^2\epsilon^2\delta]\bigr)$. When
$c$, $\Gamma$ and $(1-\rho_0)^{-1}$ are bounded independently of $N$,
this becomes
\begin{equation}
O\!\left(\frac{N^3}{\epsilon^2\delta}\right),
\end{equation}
and it remains polynomial in $N$ whenever they grow at most
polynomially. This gives a conditional PAC guarantee for the full nulling
sequence.

\section{Effect of loss}
\label{app:loss}
Theorem~\ref{thm:generic} assumes lossless operations. With loss, the
response is no longer symplectic and cannot in general be inverted, so we
instead test how well the trained mesh separates the modes of the
first-moment response. We measure the remaining mode mixing by defining
$C_{\rm off}=\|M-\operatorname{blockdiag}(M)\|_F/\|M\|_F$, where $M$ is the
response after training with Algorithm~\ref{alg:symplectic-reck} and
$\operatorname{blockdiag}(M)$ keeps only its $2\times2$ diagonal blocks.
This quantity vanishes when the modes are decoupled, regardless of
attenuation.

We use 30 targets with $N=10$ from the distribution described in
Section~\ref{sec:experiments}, with exact measurements, and compute each
nulling gate from the current lossy response. We test two loss models. To
model mode-independent loss, we attenuate every mode by the same fraction of
its power at the output. Mode separation then holds to numerical precision
for every power loss we simulate (where we simulate up to $40\%$ loss). To model mode-dependent
loss, we attenuate both output modes of every two-mode gate by the same
fraction of their power. Because different modes pass through different
numbers of gates, they accumulate unequal loss, and some mixing remains. The
mean $C_{\rm off}$ is about $2.0\%$ at $1\%$ power loss per gate and
$10.6\%$ at $5\%$ power loss per gate. Thus, the residual mixing comes from unequal loss across modes, not from
the total amount of loss. A potential way to tackle these issues from mode-dependent loss is to add dummy components so that every mode
experiences the same loss, similar to methods in Ref.~\cite{miller2015perfect}. We find in simulation that this can reduce the residual mode mixing to zero, thereby allowing full mode separation. However, the trade-off is
lower overall transmission, which falls more steeply as the loss per gate
grows. Finally, even perfect separation of the first moments cannot recover quantum
information lost to the environment or remove the noise it
adds~\cite{weedbrook2012gaussian,serafini2017quantum}.

\section{Experimental details}
\label{app:experimental-setup}

We give the target distributions, noise models and training settings used
in the numerical experiments below.

\paragraph{Targets and statistics.}
Unless stated otherwise, each data point uses 30 independent targets from the
distribution of Section~\ref{sec:experiments}, with the squeezing parameters
sorted in decreasing order. Tables and error bars report medians and 25th--75th percentiles across
trials. For Table~\ref{tab:nulling-loss}, these are taken over the pooled
local losses of all gates rather than over trials.

\subsection{Four-mode example (Fig.~\ref{fig:circuit}b)}
\label{app:four-mode}
Fig.~\ref{fig:circuit}(b) uses $N=4$, and
$r_i\sim\operatorname{Uniform}[0.3,0.4]$. Gates are trained sequentially
with automatic-differentiation gradients and noiseless responses.

\subsection{Conditioning experiment (Fig.~\ref{fig:nulling_conditioning})}
For Fig.~\ref{fig:nulling_conditioning}, we perturb each block used by the
closed-form gate as
\begin{equation}
\widehat P=P+\epsilon Z_P,
\qquad
\widehat E=E+\epsilon Z_E,
\qquad
\epsilon=10^{-8},
\label{eq:block-perturbation}
\end{equation}
where \(Z_P\) and \(Z_E\) are independent Gaussian matrices normalized to unit
Frobenius norm. The gate is computed from the perturbed blocks and applied to
the true response, and $\kappa_{\max}$ is computed from the true blocks
before each perturbation. The final single-mode gates use exact diagonal
blocks, isolating errors from the two-mode steps.

The mode-number sweep uses \(N\in\{2,5,8,11,14,17,20,25,30\}\) with 60
trials per point. The squeezing sweep uses \(N=4\) and
\(r_i\sim\operatorname{Uniform}[0,r_{\max}]\), with
$s_{\max}=20r_{\max}/\ln 10\in\{0,2,4,6,8,10\}$~dB.

For the targets approaching a nulling failure, we use $N=3$ and
\begin{equation}
S(\alpha)=e^{\alpha L}S_{\rm iso},
\qquad
S_{\rm iso}=\left(\begin{smallmatrix}
1&0&0&0&0&-1\\
0&1&0&1&0&0\\
1&0&-1&0&0&0\\
0&0&0&-1&0&0\\
0&0&0&0&0&1\\
0&1&0&1&-1&0
\end{smallmatrix}\right).
\end{equation}
At $\alpha=0$, the first nulling step has rank two and $\Delta=0$. To
build the generator $L$, we draw $S_{\rm rand}$ from the default
distribution and take $L_0$ to be the real part of the matrix logarithm of
$S_{\rm rand}S_{\rm iso}^{-1}$. We then project it as
$L=\tfrac12(L_0+J_6L_0^\top J_6)$, which ensures $L^\top J_6+J_6L=0$, so
$e^{\alpha L}$, and hence the whole path, is symplectic. The matrix
logarithm can be complex, so after taking the real part, $S(1)$ need not
equal $S_{\rm rand}$. This does not affect the sweep, since $S_{\rm rand}$
only sets a random direction away from $S_{\rm iso}$. We use
$\alpha\in\{10^{-8},10^{-7},\ldots,10^{-2}\}$, holding each trial's
generator fixed across the sweep.

\begin{figure}[t]
    \centering
    \includegraphics[width=0.85\linewidth]{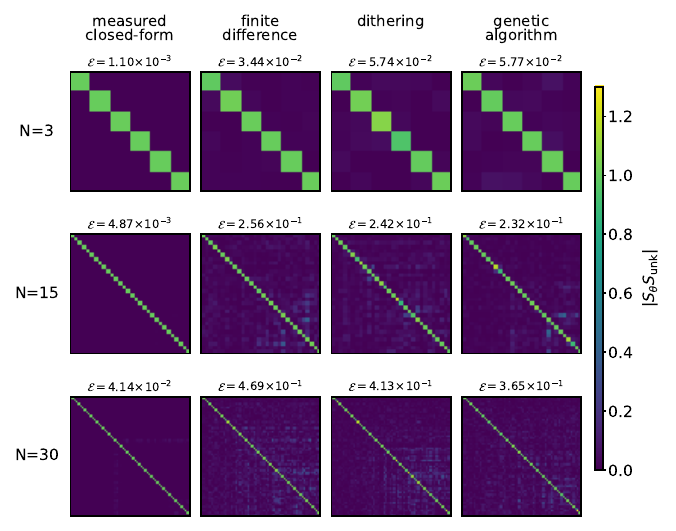}
    \caption{Response $|S_\theta S_{\rm unk}|$ after all gates in the mesh
    have been trained, for the runs in Table~\ref{tab:scaling-N} with
    $N=3$, 15 and 30 (rows) and all four methods (columns). A perfect result is
    the identity. Each panel shows the trial whose residual is closest to
    that method's median in Table~\ref{tab:scaling-N}, so different columns
    may show different targets. The residual is printed above each panel,
    and all panels share a color scale.}
    \label{fig:finite-shots-heatmaps}
\end{figure}

\subsection{Finite-shot training (Tables~\ref{tab:nulling-loss} and~\ref{tab:scaling-N}, Fig.~\ref{fig:finite-shots-heatmaps})}
\label{app:finite-shot}
For a probe displacement \(u\) and current response \(M=S_\theta S_{\rm unk}\),
each measured output mean is simulated as
\begin{equation}
\widehat{\mu}_{\ell}(u)
=(Mu)_{\ell}
+\sqrt{\frac{(MM^{\top})_{\ell\ell}}{2q}}\,
\xi_{\ell,u},
\qquad
\xi_{\ell,u}\sim\mathcal{N}(0,1),
\end{equation}
where $q$ is the number of shots per homodyne setting. This propagates the
vacuum covariance $I/2$ through $M$. We use $\eta=1$ and independent noise
for every setting. The device displacement cancels under opposite probes, so
we set it to zero.

Before the closed-form method inverts a measured diagonal block, we project
it onto $\operatorname{Sp}(2,\mathbb R)$. We take its SVD, floor the singular
values at $10^{-10}$, reverse the direction of the smallest singular value if
the determinant is negative, and normalize the determinant to one. The same
single-mode correction is used in Appendix~\ref{app:calibration-error}.

Table~\ref{tab:nulling-loss} uses $N=3$ and
$q\in\{10^2,10^3,10^4,10^5,\infty\}$. Table~\ref{tab:scaling-N} and
Fig.~\ref{fig:finite-shots-heatmaps} use $q=10^3$. Target, optimizer and
measurement-noise seeds are independent across trials. One feedback loss
evaluation uses eight homodyne settings, a closed-form two-mode gate uses
sixteen, and a closed-form single-mode gate uses eight. The feedback
optimizers use approximately 2000 loss evaluations per gate. Finite differences take 100
updates of 20 evaluations per two-mode gate and 334 updates of 6 evaluations
per single-mode gate, and the genetic algorithm runs 100 generations of
population 20. Dithering uses 100 updates of 20 evaluations per two-mode
gate and 250 updates of 8 evaluations per single-mode gate. Matching the $G=N(N+1)/2$ trained gates to the closed-form
budget gives
\begin{equation}
16q_{\rm cf}\frac{N(N-1)}{2}+8q_{\rm cf}N=2000\cdot8qG
\quad\Longrightarrow\quad
q_{\rm cf}=1000q\frac{N+1}{N}.
\label{eq:qcf-budget}
\end{equation}
Budgets are matched up to rounding of the update counts and do not
represent minimum shot requirements.

Finite differences use central differences with step \(0.08\), Adam learning
rate \(0.02\), gradient clipping at \(10\), and parameter bounds \([-5,5]\).
Dithering uses the same Adam settings, with simultaneous sinusoidal
modulation of all controls at amplitude $A=0.05$. For a window of $T$ samples,
$g_i=2(AT)^{-1}\sum_t L(t)\chi_i(t)$, where the references $\chi_i$ are sine and cosine
pairs at odd harmonics. Two-mode gates use harmonics $1,3,5,7,9$ with $T=20$,
and single-mode gates use the first three references with $T=8$.
Each sample uses fresh homodyne noise. The amplitude was selected on held-out
targets starting at seed 500, with the finite-difference step checked on the
same targets. These settings remain fixed in the reported sweep.
The genetic algorithm uses elite fraction \(0.3\), parent fraction \(0.5\),
crossover by random per-parameter averaging of two parents, mutation scale
\(0.04\), mutation decay \(0.995\), and the same bounds. These
hyperparameters are fixed across all runs.

\subsection{Fixed calibration errors (Fig.~\ref{fig:insitu_vs_tomography_calibration})}
\label{app:calibration-error}

The finite-shot experiments above assume exact gate implementation. Here
we introduce fixed gate errors to compare in-situ training with tomography
followed by offline compilation.
Fig.~\ref{fig:insitu_vs_tomography_calibration} uses $N=10$, 30 targets,
and 13 error bounds $\epsilon_{\max}=10^{-3+k/6}$, $k=0,\ldots,12$. Every
gate $G_g$ is implemented as
\begin{equation}
\widetilde G_g=E_g(\boldsymbol\delta_g)G_g,
\qquad
\delta_{g,a}\overset{\mathrm{iid}}{\sim}
\operatorname{Uniform}[-\epsilon_{\max},\epsilon_{\max}].
\label{eq:fixed-error-gate}
\end{equation}
For a two-mode gate, the ten parameters enter the Bloch--Messiah form
\begin{equation}
E_g=O(U_{g,1})\,
\operatorname{diag}(e^{\delta_{g,5}},e^{-\delta_{g,5}},
                    e^{\delta_{g,6}},e^{-\delta_{g,6}})
\,O(U_{g,2}),
\end{equation}
with $O(U)$ from Eq.~\eqref{eq:orthosymp},
$U_{g,1}=U_{\rm err}(\delta_{g,1},\ldots,\delta_{g,4})$,
$U_{g,2}=U_{\rm err}(\delta_{g,7},\ldots,\delta_{g,10})$, and
\begin{equation}
U_{\rm err}(a,b,\theta,\phi)
=\operatorname{diag}(1,e^{i\phi})
\begin{pmatrix}\cos\theta&-\sin\theta\\
\sin\theta&\cos\theta\end{pmatrix}
\operatorname{diag}(e^{ia},e^{ib}).
\end{equation}
For a single-mode gate,
$E_g=\mathcal R(\delta_{g,1})
\operatorname{diag}(e^{\delta_{g,2}},e^{-\delta_{g,2}})
\mathcal R(\delta_{g,3})$. Angles are in radians and squeezing parameters
are dimensionless, so a single bound applies to both. Note that
$\epsilon_{\max}$ bounds each error parameter, not the size of the resulting
matrix error. The errors are drawn once per trial and shared by all four
curves. Across the sweep, the same draws from
$\operatorname{Uniform}[-1,1]$ are rescaled by $\epsilon_{\max}$.

The tomography baseline estimates every entry of $S_{\rm unk}$ once with
$\pm$ probes and projects the estimate $\widehat S$ onto
$\operatorname{Sp}(2N,\mathbb R)$,
\begin{equation}
S_{\rm proj}=\widehat S
\left(J^\top\widehat S^\top J\widehat S\right)^{-1/2},
\label{eq:symplectic-projection}
\end{equation}
using the real principal inverse square root. It then computes every gate offline from this model. After each gate, it
updates the model assuming the gate was implemented exactly, so errors in
earlier gates are not accounted for when later gates are computed. In-situ training instead measures the two blocks needed by
each two-mode gate, and each diagonal block for the final layer, on the
physical device after all earlier imperfect gates have acted.

Both methods use the minimum-squeezing gate and the single-mode correction of
Appendix~\ref{app:finite-shot}. In-situ training uses
$16N(N-1)/2+8N=8N^2$ homodyne settings, the same as full tomography, and
both use $q=10^4$ shots per setting. The noiseless curves keep the same gate
errors but use exact responses. All 30 trials are valid at every
$\epsilon_{\max}$. The right panel reports quartiles of
$\mathcal E_{\rm tomog}-\mathcal E_{\rm in\text{-}situ}$ computed within each
paired trial.

\subsection{Computational resources}
All simulations can run on a standard CPU. Closed-form training takes less than a second per target for every size we
test. The feedback optimizers take longer, and the full set of
finite-shot experiments took approximately two hours in total using
16 parallel CPU workers.\end{document}